\documentclass[aps,showpacs,pra]{revtex4}

\usepackage[tbtags]{amsmath}
\usepackage{amssymb}
\usepackage{amsfonts}
\usepackage[dvips]{graphicx}

\begin{document}
\title{Time-resolved measurement of photon states using two-photon interference with photons from short-time reference pulses}
\author{Changliang Ren}
\author{Holger F. Hofmann}
\email{hofmann@hiroshima-u.ac.jp}
\affiliation{
Graduate School of Advanced Sciences of Matter, Hiroshima University,
Kagamiyama 1-3-1, Higashi Hiroshima 739-8530, Japan}
\affiliation{JST, CREST, Sanbancho 5, Chiyoda-ku, Tokyo 102-0075, Japan}

\begin{abstract}
To fully utilize the energy-time degree of freedom of photons for optical quantum information processes, it is necessary to control and characterize the quantum states of the photons at extremely short time scales. For measurements beyond the time resolution of available detectors, two-photon interference with a photon in a short time reference pulse may be a viable alternative. In this paper, we derive the temporal measurement operators for the bunching statistics of a single photon input state with a reference photon. It is shown that the effects of the pulse shape of the reference pulse can be expressed in terms of a spectral filter selecting the bandwidth within which the measurement can be treated as an ideal projection on eigenstates of time. For full quantum tomography, temporal coherence can be determined by using superpositions of reference pulses at two different times. Moreover, energy-time entanglement can be evaluated based on the two-by-two entanglement observed in the coherences between pairs of detection times.
\end{abstract}

\pacs{
03.65.Wj,   
42.50.Dv,   
42.65.Re,   
03.67.Mn    
}

\maketitle

\section{Introduction}

Photons are one of the most versatile systems for the implementation of quantum information processes. The most accessible degrees of freedom are polarization and the spatial degrees of freedom defined by optical paths. In addition, it is possible to encode quantum information in the different frequency modes along the direction of propagation. Encoding information in this energy-time degree of freedom may be more robust against mechanical instabilities than the alternatives and could therefore be especially useful in long-distance quantum communication \cite{Zbinden, Tittel}.

Energy-time entanglement was first demonstrated by Franson \cite{Franson}, and has been applied in quantum cryptography \cite{Tittel1, Ribordy} and quantum communication \cite{Brendel, Marcikic, Marcikic1}. Originally, these approaches are based on comparatively long timescales, so that the necessary time resolution can be achieved by the photon detection systems directly. However, recent technological developments make it possible to generate entanglement on time scales much shorter than the time resolution of available detectors \cite{Harris,Nasr,Hendrych}. To fully utilize the potential of such broadband sources of entangled photons, it is necessary to develop measurement systems that can characterize the quantum states of photons on extremely short time scales.

In previous experimental work, short time correlations between entangled photons were confirmed by observing the rate of two photon absorption in second harmonic generation, effectively reversing the role of the source to serve as a detector \cite{Sensarn,ODonnell}. However, the measurement information so obtained is insufficient for a complete characterization of the quantum state. More detailed information can be obtained by the two-photon interference between a pair of down-converted photons, where the shape of the Hong-Ou-Mandel dip can provide spectral information about the photon pairs \cite{HOM1,Okamoto}. However, this method only demonstrates that the photons have some time dependent correlation, without deciding whether this correlation is between detection times or represent a time dependence of energy, e.g. in chirped pulses.

An independent time standard can be realized by using reference pulses from a separate photon source. With such independently generated references, it is possible to use the photon bunching effect of the two photon interference between one reference photon and the signal photon to achieve a time resolved measurement at the time scale determined by the pulse time of the reference pulse. Intuitively, bunching should only occur when the reference photon and the input photon coincide in time. However, bunching is also sensitive to the temporal coherence of the reference pulse. To optimize the information gained about the state of the input photon, it is essential to understand these effects of temporal coherences in the reference pulses on the bunching statistics. In the following, we therefore analyze the quantum statistics obtained from appropriate short time reference pulses and explore the possibilities of quantum tomography and entanglement verification in the time domain based on two-photon interference.

In Sec. \ref{sec:bunch}, we describe the coincidence rates in the
output ports of the two-photon interference in terms of a projective
measurement operator defined by the optical quantum state of the
photon from the reference pulse. In Sec. \ref{sec:delay}, the
effects of time delays used to scan the temporal features of the
signal photon state are considered. Significantly, the measurement
is equivalent to the projection on an ideal eigenstate of time if
the bandwidth of the input state is narrower than that of the
reference pulse. It is therefore possible to optimize the reference
pulse shape according to the basic optical properties of the input.
In Sec. \ref{sec:cohere}, we show how a superposition of reference
pulses at two times can be used to obtain the temporal coherence of
the input state. We can then achieve full quantum tomography in the
time domain. In Sec. \ref{sec:entangle}, we apply the results to
derive an experimental criterion for the verification of the
temporal entanglement of down-converted photon pairs.

\section{Two-photon interference with a reference photon}
\label{sec:bunch}

Photon bunching of the Hong-Ou-Mandel type is observed when two photons are incident on a $50:50$
beam splitter from opposite sides. If the photons share the same temporal and transverse coherence, two photon interference eliminates the possibility that they will exit the beam splitter on opposite sides. However, any mismatch in the temporal or transverse coherence will result in the appearance of photons exiting at opposite sides. Coincidence detection of photons on opposite sides of the beam splitter is therefore a direct measure of the mismatch between the quantum states of the photons that entered the beam splitter.

\begin{figure}
[ht]
\begin{center}
\includegraphics[width=0.3\textwidth]{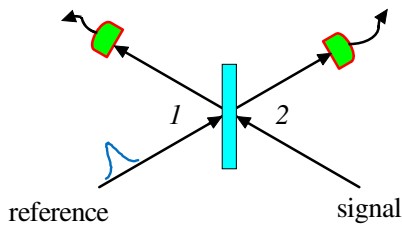}
\caption{\label{fig1} Illustration of a bunching measurement of an unknown photon state using a short time reference photon and a $50:50$ beam splitter.}%
\end{center}
\end{figure}

To obtain temporal features of a signal photon, we consider the use of a reference photon from a short time pulse, as illustrated in Fig. \ref{fig1}. The transformation of the light field at the beam splitter can be represented by the transformation of the temporal annihilation operators,
\begin{eqnarray} \label{eq:trans}
\hat{b}_{1}(t)&=&\frac{1}{\sqrt{2}}(\hat{a}_{1}(t)+\hat{a}_{2}(t))
\nonumber \\
\hat{b}_{2}(t)&=&\frac{1}{\sqrt{2}}(\hat{a}_{1}(t)-\hat{a}_{2}(t)),
\end{eqnarray}
where $\hat{a}_{1}(t)$ and $\hat{a}_{2}(t)$ are the input field
operators and $\hat{b}_{1}(t)$ and $\hat{b}_{2}(t)$ are the output
field operators for a given input time $t$. Since the photon in
input port $1$ is the reference photon, its input state is known and
given by the reference state $\mid \Phi_{\mathrm{ref}}\rangle$. In
the single photon time basis, this state is represented by
\begin{equation} \label{refer}
\mid  \Phi_{\mathrm{ref}}\rangle=\int
\phi(t)\hat{a}_{1}^{\dag}(t)\mathrm{d}t\mid 0\rangle,
\end{equation}
where $\phi(t) = \langle t\mid \Phi_{\mathrm{ref}}\rangle$ is
the time dependence of the reference pulse amplitude.

In general, the probability of a joint detection of one
photon in output $1$ at time $t_{1}$ and one photon in outport $2$
at time $t_{2}$ is
\begin{equation} \label{Coin}
G^{(2)}(t_{1},t_{2})=\langle \hat{b}_{1}^{\dag}(t_{1})\hat{b}_{2}^{\dag
}(t_{2})\hat{b}_{2}(t_{2})\hat{b}_{1}(t_{1})\rangle.
\end{equation}
We can transform this expression into an expectation value for the coherences of the two photon input state using Eq. (\ref{eq:trans}). Since the reference state is known, the joint probability can be expressed in terms of a projection on the single photon state in the signal port $2$. The projection operator for detections at times $t_1$ and $t_2$ can then be given by
\begin{equation}\label{ms}
\hat{\Pi}_{s}(t_{1},
t_{2})=\frac{1}{4}\left(\phi(t_{1})\phi^{*}(t_{1})\mid t_{2}\rangle
\langle t_{2}\mid-\phi(t_{1})\phi^{*}(t_{2})\mid t_{2}\rangle
\langle t_{1}\mid\\-\phi(t_{2})\phi^{*}(t_{1})\mid t_{1}\rangle
\langle t_{2}\mid+\phi(t_{2})\phi^{*}(t_{2})\mid t_{1}\rangle
\langle t_{1}\mid\right)
\end{equation}
However, this measurement operator assumes perfect time resolution of the detection. In realistic photon detectors, the detection times that can be resolved are usually much longer than the pulse times of the available reference pulses. We can therefore assume that the photon time will be completely unknown, so that the actual measurement operator of the photon coincidence is obtained by integrating over all measurement times,
\begin{equation}\label{eq:MS}
\hat{M}_{s}=\int \int
\hat{\Pi}_{s}(t_{1}, t_{2})
\mathrm{d}t_{1}\mathrm{d}t_{2}=\frac{1}{2}-\frac{1}{2}\mid
\Phi_{\mathrm{ref}}\rangle \langle \Phi_{\mathrm{ref}}\mid.
\end{equation}
Thus, the effect of two-photon interference with the reference photon on the coincidence counts is given by a single photon measurement operator defined by the projection on the state of the reference. In the following, we will consider the effects of the spectral and temporal features of the reference on the time resolution of the measurement.

\section{Time resolved measurement and bandwidth limitation}
\label{sec:delay}

Time resolved measurements can be realized by scanning the peak time of the reference pulse using an appropriate time-delay. To represent this time delay in our formalism, we can simply shift the temporal wavefunction, so that the amplitude at input time $t^\prime$ for a delay time of $t$ corresponds to
\begin{equation}
\label{eq:timeshift}
\langle t^{'}\mid \Phi(t)\rangle=\phi(t^{'}-t)
\end{equation}
Alternatively, this time shift can be expressed by a phase factor in the frequency representation,
\begin{equation} \label{eq:phasefactor}
\langle \omega\mid \Phi(t)\rangle=e^{i\omega t}\langle \omega\mid
\Phi(0)\rangle.
\end{equation}
An ideal measurement of time would project on the state $\mid t \rangle$ defined by a delta function in time or, in the frequency representation, by
\begin{equation} \label{eq:ideal}
\langle \omega\mid t \rangle= \frac{1}{\sqrt{2 \pi}} e^{i\omega t}.
\end{equation}
The comparison of real state (\ref{eq:phasefactor}) and ideal state (\ref{eq:ideal}) shows that the difference can be expressed by a time-shift independent attenuation of the frequency components, so that
\begin{equation} \label{eq:filter}
\mid \Phi(t)\rangle = \left(\int \sqrt{2 \pi} \,\langle \omega\mid \Phi(0)\rangle
\;\;\mid \omega\rangle\langle \omega\mid
\mathrm{d}\omega \right)
\mid t\rangle
= \hat{F} \; \mid t\rangle.
\end{equation}
The problem of time resolution can therefore be expressed completely in terms of the filter operation $\hat{F}$ that converts the ideal time eigenstate into the reference pulse. The measurement operator for a reference time $t$ is then given by
\begin{equation} \label{eq:timeoperator}
\hat{M_{s}}(t)=\frac{1}{2}-\frac{1}{2}\hat{F}\mid t\rangle \langle
t\mid\hat{F}^\dagger.
\end{equation}
To determine the overlap between $\mid \Phi(t)\rangle$ and an unknown input state $\mid \psi \rangle$, we can first apply the adjoint filter operator $\hat{F}^\dagger$ to modify the frequency amplitudes of the input state. The overlap of the original state with the reference state is then equal to the overlap of the modified state with an eigenstate of time.

In general, the eigenvalues of $\hat{F}^\dagger$ are equal to $\sqrt{2 \pi}$ times the complex conjugate of the frequency amplitude $ \langle \omega \mid \Phi(0)\rangle$ of the reference pulse. Since dispersion effects caused by phase changes in the frequency components should be avoided, it is best to use transform limited reference pulses with real and positive amplitudes. In this case, $\hat{F}$ is a self-adjoint operator acting like a dispersion-free band-pass filter. Optimally, the frequency spectrum of the reference pulse should be approximately rectangular, resulting in constant transmission of a specific bandwidth. The temporal measurement will then be completely accurate for all input fields within that bandwidth.

It may also be worth noting that the depth of the dip in the coincidence count rate described by the measurement operator in Eq.(\ref{eq:timeoperator}) depends on the bandwidth $\Delta \omega$ of the reference pulse. For a rectangular spectrum of the reference pulse, the average number of coincidence counts $N_c$ from an input state $\mid \psi \rangle$ with a frequency spectrum entirely within the bandwidth of $\Delta \omega$ is given by
\begin{eqnarray} \label{eq:coincidence}
N_{c}=\frac{1}{2}-\frac{\pi}{\Delta \omega} |\langle t \mid \psi \rangle|^2.
\end{eqnarray}
To optimize the signal-to-noise ratio of the measurement, it is therefore desirable to use the minimal bandwidth acceptable for the respective input states.

\section{Measurement of temporal coherence}
\label{sec:cohere}

As shown in the previous section, the probability distribution of photons in time can be obtained from the photon bunching characteristics of two-photon interference with photons in a single short-time reference pulse. In terms of the temporal density matrix, we can then reconstruct the diagonal elements $\langle t \mid \hat{\rho} \mid t \rangle$. A complete tomographic reconstruction of an unknown quantum state can be achieved by determining all elements $\langle t_1 \mid \hat{\rho} \mid t_2 \rangle$, including the off-diagonal elements that describe coherences between two different times $t_1$ and $t_2$.

\begin{figure}
[ht]
\begin{center}
\includegraphics[width=0.33\textwidth]{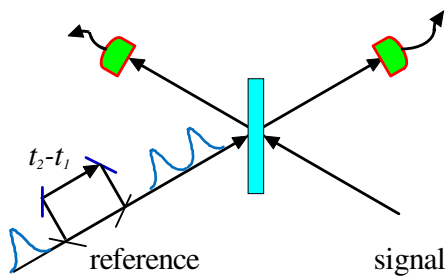}
\caption{\label{fig2} Setup for the measurement of temporal coherence using coherent
superpositions of reference pulses at two different times, $t_{1}$ and $t_{2}$.}%
\end{center}
\end{figure}

Since photon bunching with a reference corresponds to a projection on the state of this reference, coherences between two times can be probed by using superpositions of two short-time pulses as a reference. The setup for this kind of measurement is shown in Fig. \ref{fig2}. The coherent reference is generated by interference between the original pulse and a time-delayed pulse, so that the reference is in an equal superposition of input times $t_1$ and $t_2$,
\begin{equation} \label{eq:superpos}
\mid \Phi_{\mathrm{ref}}\rangle
=\frac{1}{\sqrt{2}} \hat{F} (\mid t_1 \rangle+\mid t_2 \rangle),
\end{equation}
where we used Eq. (\ref{eq:filter}) to express the superposition in terms of time eigenstates and the filter operator $\hat{F}$ representing the permitted bandwidth.

Unfortunately, the phase in this superposition is fixed to zero, because the shape of the initial reference pulse defines the precise phase at peak time $t_i$. To achieve complete quantum tomography, we need to consider the possibility of obtaining phase information from the dependence of coherence on the time difference between $t_1$ and $t_2$. Since our measurement has a limited bandwidth centered around a carrier frequency of $\omega_0$, we can assume that small shifts in the time difference do result in a corresponding phase shift $\varphi = \Delta t/\omega_0$. Although the time eigenstates $\mid t\rangle$ and $\mid t+\frac{\varphi}{\omega_{0}}\rangle$ are orthogonal, they are nearly indistinguishable within the bandwidth considered, as long as $\varphi$ is sufficiently smaller than $\omega_0/\Delta \omega$. Specifically, the application of the filter function $\hat{F}$ ensures that
\begin{equation}
\hat{F}\mid t+\frac{\varphi}{\omega_{0}}\rangle\approx
e^{-i\varphi}\hat{F}\mid  t\rangle.
\end{equation}
Arbitrary phase shifts can then be realized with $-\pi \leq \varphi \leq \pi$. In this range, it is usually reasonable to consider the above approximation as valid, so that the state $\mid
t+\frac{\varphi}{\omega_{0}}\rangle$ can be replaced by $e^{-i\varphi}\mid t\rangle$. A quantitative test of this assumption can be obtained from the normalized overlap $\sigma$ of the filtered states,
\begin{eqnarray} \label{eq:fidelity}
\sigma = \frac{|\langle t\mid\hat{F}^{\dagger} \hat{F}\mid
t+\frac{\varphi}{\omega_{0}}\rangle|^{2}}
{\langle t\mid\hat{F}^{\dagger} \hat{F} \mid t\rangle \;
\langle t+\frac{\varphi}{\omega_{0}}\mid\hat{F}^{\dagger} \hat{F}\mid
t+\frac{\varphi}{\omega_{0}}\rangle}.
\end{eqnarray}
In most cases, $\sigma$ will be close to one for the whole range of phase shifts. It is then safe to assume that we can generate superpositions of different times with arbitrary phases. The measurement operators obtained with these superpositions read
\begin{equation} \label{eq:mscohere}
\hat{M_{s}}(t_{1}, t_{2},\varphi)=\frac{1}{2}-\frac{1}{4}\hat{F}\, \left(
\mid t_{1}\rangle\langle t_{1}\mid + e^{-i\varphi} \mid t_{1}\rangle\langle t_{2}\mid
+ e^{i\varphi} \mid t_{2}\rangle\langle t_{1}\mid + \mid t_{2}\rangle\langle t_{2}\mid
\right)\,\hat{F}^\dagger.
\end{equation}
By scanning $t_1$, $t_2$ and $\varphi$ over the appropriate range of
values, a complete tomographic reconstruction of the density matrix
$\langle t_1 \mid \hat{\rho} \mid t_2 \rangle$ of the input photons
is possible.

\section{Verification of energy-time entanglement}
\label{sec:entangle}

\begin{figure}
[ht]
\begin{center}
\includegraphics[width=0.45\textwidth]{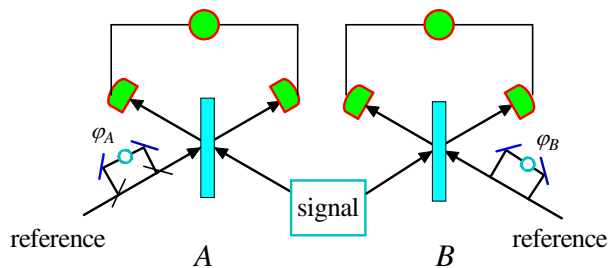}
\caption{\label{fig3} Setup for correlated measurements on energy-time entangled photon pairs.}%
\end{center}
\end{figure}

In principle, it is a straightforward matter to apply the quantum tomography scheme described in the previous section to photon pairs in energy-time entangled states. As shown in Fig. \ref{fig3}, it is merely necessary to extend the two photon coincidence measurements for a signal photon and a reference photon to the four photon coincidences observed for two signal photons with two separate references. The measurement operator of the four-fold coincidence then reads
\begin{equation}
\label{eq:fourfold}
\hat{M}_p(t_{1A},t_{2A},\varphi_A; t_{1B}, t_{2B}, \varphi_B) =
\hat{M}_s(t_{1A},t_{2A},\varphi_A) \otimes \hat{M}_s(t_{1B}, t_{2B}, \varphi_B),
\end{equation}
where the operators $\hat{M}_s$ describe the single photon results according to Eq.(\ref{eq:mscohere}). This operator can be further separated into a positive background, negative contributions from local bunching at either $A$ or $B$, and a positive term from the correlation between bunching effects. Since only the last term is required for the characterization of entanglement, we can focus on this term, given by the projection on the product of the two references in $A$ and in $B$,
\begin{equation}
\label{eq:tworef}
\mid \Phi_{A}; \Phi_B \rangle = \frac{1}{2} \hat{F} \otimes \hat{F} \left(
\mid t_{1A}, t_{1B} \rangle + e^{i\varphi_A} \mid t_{2A}, t_{1B}\rangle
+ e^{i\varphi_B} \mid t_{1A}, t_{2B}\rangle
+ e^{i (\varphi_A +\varphi_B)} \mid t_{2A}, t_{2B}\rangle
\right)
\end{equation}
The projector corresponding to this reference state can be represented in the four dimensional subspace spanned by the temporal eigenstates of times $t_{1A}, t_{2A}$ in path $A$, and $t_{1B}, t_{2B}$ in path $B$. In  this two-by-two Hilbert space, the projector can be given by the matrix
\begin{equation}\label{eq:matrix}
\hat{P}(\varphi_{A},
\varphi_{B})=\frac{1}{4}\left(\begin{array}{cccc}
 1&e^{-i\varphi_{B}}&e^{-i\varphi_{A}}&e^{-i(\varphi_{A}+\varphi_{B})}\\
e^{i\varphi_{B}}&1&e^{-i(\varphi_{A}-\varphi_{B})}&e^{-i\varphi_{A}}\\
 e^{i\varphi_{A}}&e^{i(\varphi_{A}-\varphi_{B})}&1&e^{-i\varphi_{B}}\\
e^{i(\varphi_{A}+\varphi_{B})}&e^{i\varphi_{A}}&e^{i\varphi_{B}}&1\\
\end{array}\right),
\end{equation}
where the basis states are $\{\mid t_{1A}, t_{1B}\rangle, \mid
t_{1A}, t_{2B}\rangle, \mid t_{2A}, t_{1B}\rangle, \mid t_{2A},
t_{2B}\rangle\}$. By varying the phases in $A$ and in $B$, it is
possible to distinguish contributions with different phase factors.
In particular, it is possible to identify the coherence between
$\mid t_{1A},t_{1B}\rangle$ and $\mid t_{2A},t_{2B}\rangle$ from the
coincidence data, since it is the only term that depends on the sum
of $\varphi_A$ and $\varphi_B$. This term describes a genuine
two-photon coherence and is particularly useful for the verification
of temporal entanglement.

For single photon states, the coherence between $\mid t_1 \rangle$ and $\mid t_2 \rangle$ is limited by $|\langle t_1 \mid \hat{\rho} \mid t_2 \rangle| \leq 1/2$. Therefore, the two photon coherence between
$\mid t_{1A},t_{1B}\rangle$ and $\mid t_{2A},t_{2B}\rangle$ achieved by separable states is limited to
\begin{equation} \label{eq:sep}
|\langle t_{1A}, t_{1B}\mid \hat{\rho}_{\mathrm{sep.}}\mid t_{2A},
t_{2B}\rangle| \leq \frac{1}{4}.
\end{equation}
For entangled states, the same coherence can be as high as $1/2$. It is therefore possible to verify temporal entanglement directly by detecting values of $\langle t_{1A}, t_{1B}\mid \hat{\rho}\mid t_{2A},t_{2B}\rangle>1/4$ at a fixed set of detection times. In typical experiments using parametric down-conversion, photon pairs will be emitted simultaneously, so entanglement can be verified for $t_{1A}=t_{1B}$ and $t_{2A}=t_{2B}$. For time differences of $t_2-t_1$ smaller than the correlation time of the down-conversion process, single photon cohernces will
emerge between $t_1$ and $t_2$, resulting in a reduction of the observed entanglement. This transition between two-photon coherence and single photon coherences can be used to identify the timescale of the entanglement. For instance, it may be reasonable to define the timescale achieved by broadband entanglement sources as the time difference $t_2-t_1$ at which the coherence $\langle t_{1}, t_{1}\mid \hat{\rho}\mid t_{2},t_{2}\rangle$ crosses the value of $3/8$ halfway between maximal entanglement and separability. Thus, bunching with coherent superpositions of short-time reference pulses provides an experimental method of identifying the ultra-short quantum correlation times of broadband entanglement sources.

\section{Conclusions}

We have shown that the quantum state of an unknown input photon can be completely determined by two-photon interference with reference photons from short-time references pulses. In particular, the bunching effect observed with a single reference pulse corresponds to a projection on a time eigenstate, where the limited time resolution can be represented by a filter function that eliminates frequencies outside of the bandwidth of the reference pulse. To obtain an optimal signal-to-noise ratio, it is desirable to make this bandwidth as narrow as possible without losing too much of the input signal. Two-time coherences can be evaluated using superpositions of two pulses shifted to appropriate times. The phase relation between these pulses can be controlled in the conventional manner, by time shifts shorter than one period of the central frequency of the reference pulse. It is then possible to achieve a complete tomographic reconstruction of the temporal density matrix from the two-photon interference data.

The application of our tomography scheme to energy-time entanglement is straightforward. However, it may often be sufficient to evaluate the entanglement from only a few selected measurements. In the present scheme, the effective two-level entanglement between two pairs of detection times can be determined from the phase dependence of the four photon coincidences in the output ports. It is then possible to characterize the time dependence of entanglement in terms of the two-photon coherences between two different detection times, where the characteristic timescale might be given in terms of the time separation where the coherence is halfway between its maximal and its separable value. The entanglement times achieved by various sources of energy-time entangled photon pairs can then be determined experimentally using well-defined criteria.

In conclusion, photon bunching with short time references may prove
to be a helpful addition to the available experimental methods for
the study of energy-time entanglement and related quantum effects on
ultrashort timescales. Our results show that the data obtained in
such measurements provides a direct and intuitive image of temporal
quantum coherence, permitting the definition of clear and accessible
criteria for the temporal features of photon states. The present
work may thus lay the foundations for detailed investigations of the
energy-time degrees of freedom in multi-photon quantum statistics.

\section*{Acknowledgment}
Part of this work has been supported by the Grant-in-Aid program of the Japanese Society for the Promotion of Science, JSPS.


\end{document}